\documentclass[twocolumn]{revtex4}
\usepackage{amssymb}
\usepackage{graphicx}
\usepackage{dcolumn}
\usepackage{bm}
\usepackage{amsmath}
\usepackage{epsfig}
\usepackage{braket}
\usepackage[paperwidth=210mm,paperheight=297mm,centering,hmargin=2cm,vmargin=2.5cm]{geometry}

\begin{document}

\title{Atoms and Photons - Their Interaction Dynamics}

\author{Max F. Frenzel}
\affiliation{Blackett Laboratory, Imperial College London, Prince Consort Rd, London SW7 2BW, UK}
\date{April 30, 2012}

\begin{abstract}

The Jaynes-Cummings model (JCM) describes the interaction of a two-level atom with a single quantised field mode in an optical cavity. In this BSc project we derive the basic Rabi model Hamiltonian and show how it leads to the JCM Hamiltonian. The results are then used to analyse the general time evolution of a bipartite atom-field system and its properties, such as entanglement. We extend a scheme developed by Browne and Plenio \citep{Browne2002a} to entangle two cavities that interact with an atom. Our scheme allows to produce states that are arbitrarily close to a Bell state with a success probability greater than 0.75, whereas the success rate of the original method tends to 0 for perfect fidelity. Further, we extend both schemes by allowing the atom or the cavity to be in a thermal state and find that the fidelity of the produced states is reduced with increasing temperature. Finally, we apply our scheme to quantum teleportation, where we find that the fidelity of the teleported state compared to the original state also decreases as the temperature is increased.

\end{abstract}

\maketitle

\section{Introduction}
The Jaynes-Cummings model (JCM) is of particular theoretical interest  since it provides the simplest fully quantum mechanical description of the interaction between a two-level atom and a quantised field \citep{Jaynes1963a,Boukobza2005}. In many cases, the full evolution of such a bipartite system can be solved analytically. Hence the JCM represents a unique tool for analytically studying many properties of these systems that can otherwise only be evaluated numerically. Of particular interest is the entanglement that can be generated between the atom and the field, or even between individual field modes that interact with the atom, as shall be shown later.\\
In this project we focus on deriving the JCM and applying it to study the time evolution of the bipartite system, starting from various initial conditions. In section II we will derive the Rabi model Hamiltonian and show how, under the rotating wave approximation (RWA), it leads to the JCM Hamiltonian. We will use effective Hamiltonian theory \citep{James2007} to justify the RWA and show how the Hamiltonian behaves in the near resonant and far off-resonant limits. Section III will utilise the found Hamiltonians to develop evolution operators that describe the full dynamics of the system under the interaction. We also look at some particular examples with simple initial conditions that give a general feeling for the behaviour of the system as it evolves.\\
In Section IV we make use of all the tools we developed to study a particular system that was initially proposed by Browne and Plenio \citep{Browne2002a}, which consists of an atom that is sent through two optical cavities and briefly interacts with each cavity mode, thereby entangling the two cavities. We propose an improvement on the original scheme and look at the effect of thermal interaction between the system and the environment, by allowing either the atom or the cavities to be in thermal states. Finally, in section V we evaluate if the generated state could be used for quantum teleportation and how the fidelity of the teleported state depends on temperature.

\section{Derivation of the Hamiltonian}

Starting from the Hamiltonian of an electron which is bound to a nucleus under an external field \cite{Gerry2004},
\begin{equation} \label{eq:Hamiltonian}
\hat{H}(\mathbf{r},t) = \frac{1}{2m}[\mathbf{\hat{P}} + e\mathbf{A}(\mathbf{r},t)]^2 - e\Phi(\mathbf{r},t) + V(r),
\end{equation}
where $\mathbf{A}(\mathbf{r},t)$ and $\Phi(\mathbf{r},t)$ are the vector and scalar potentials of the external field and $V(r)$ is the  Coulomb potential that binds the electron to the nucleus, we used gauge transformations, particularly the Coulomb gauge, to show that the Hamiltonian can be rewritten as 
\begin{equation} \label{eq:HamiltonianTrans}
\hat{H}'(\mathbf{r},t) = \hat{H}_0(\mathbf{r},t) - \mathbf{\hat{d}}\cdot\mathbf{\hat{E}}(t),
\end{equation}
where $\hat{H}_0(\mathbf{r},t)$ is the Hamiltonian \eqref{eq:Hamiltonian} in the absence of an external field, $\mathbf{\hat{d}} = -e\mathbf{r}$ is the dipole operator, and $\mathbf{\hat{E}}(t)$ is the operator representing the time varying electric field \cite{Gerry2004, Bauch2011,Nietner2010}. The interaction between atom and field is represented by the term $\hat{H}_{int} = -\mathbf{\hat{d}}\cdot\mathbf{\hat{E}}(t)$. By considering the matrix elements of this operator in the atomic basis $\ket{e}$ and $\ket{g}$, the excited and ground state respectively, one can show that (setting $\hbar=1$, as shall be done throughout the rest of this discussion)
\begin{equation} \label{eq:HamiltonianInt}
\hat{H}_{int} = \lambda(\hat{\sigma}_+ + \hat{\sigma}_-)(\hat{a} + \hat{a}^{\dagger}),
\end{equation}
where $\lambda$ is the coupling strength, depending on the dipole moment and the electric field, $\sigma_+ = \ket{e}\bra{g}$ and $\sigma_- = \ket{g}\bra{e}$ are the atomic transition operators, and the field raising and lowering operators  $\hat{a}^{\dagger}$ and $ \hat{a}$ have their usual meaning.\\
$\hat{H}_0$ can be found by considering the energy of the atom and the field individually. By taking the zero of the atomic energy midway between the two levels, the atomic part of the Hamiltonian is simply $\hat{H}_A = \frac{1}{2}\omega_0\hat{\sigma}_3$, where $\omega_0$ is the energy difference between the atomic levels and $\hat{\sigma}_3 = \ket{e}\bra{e}-\ket{g}\bra{g}$ is the atomic inversion operator \cite{Gerry2004}. The Hamiltonian for the field takes an equally simple form of $\hat{H}_F = \omega(\hat{a}^{\dagger}\hat{a}+\frac{1}{2})$, with $\omega$ being the frequency of the radiation \cite{Gerry2004}.\\
Combining the individual terms and dropping the vacuum energy contribution in the field Hamiltonian (since it does not contribute to the dynamics), gives the Rabi model Hamiltonian
\begin{equation} \label{eq:Rabi}
\hat{H} = \frac{1}{2}\omega_0\hat{\sigma}_3 + \omega\hat{a}^{\dagger}\hat{a} + \lambda(\hat{\sigma}_+ + \hat{\sigma}_-)(\hat{a} + \hat{a}^{\dagger}).
\end{equation}
Under certain conditions the interaction term can be further simplified. Using the approach described by James and Jerke \cite{James2007} for deriving an effective interaction Hamiltonian, we showed that near resonance, i.e. for small detuning $\Delta = \omega_0 - \omega \ll \omega$, the interaction Hamiltonian \cite{Barnett2003} is given by
\begin{equation} \label{eq:JCM_int}
\hat{H}_{I} = \Delta\frac{\hat{\sigma}_3}{2} + \lambda\bigl(\hat{\sigma}_+\hat{a} + \hat{\sigma}_-\hat{a}^{\dagger} + \mathcal{O}(\frac{\lambda}{\omega})\bigr).
\end{equation}
Hence if we limit our focus to systems for which $\omega \gg \Delta , \lambda$ we can neglect the $\mathcal{O}(\frac{\lambda}{\omega})$ term in \eqref{eq:JCM_int}, having justified the rotating wave approximation that is often made in the literature \cite{Boukobza2005,Phoenix1991b,Gerry2004}, to arrive at the full JCM Hamiltonian for the near resonant case,
\begin{equation} \label{eq:JCM}
\hat{H} = (\omega + \Delta)\frac{\hat{\sigma}_3}{2} + \omega\hat{a}^{\dagger}\hat{a} + \lambda(\hat{\sigma}_+\hat{a} + \hat{\sigma}_-\hat{a}^{\dagger}).
\end{equation}
A similar treatment for the case of large detuning, where $\omega \gg \Delta \gg \lambda$, leads to the interaction picture Hamiltonian
\begin{equation} \label{eq:LargeDet_eff}
\hat{H}_{I} = \frac{\lambda^2}{\Delta}\Bigl(\hat{\sigma}_3\hat{a}^{\dagger}\hat{a} + \ket{e}\bra{e}\Bigr),
\end{equation}
which gives the full JCM Hamiltonian for the off-resonant case,
\begin{equation} \label{eq:LargeDet}
\hat{H} = (\omega + \Delta)\frac{\hat{\sigma}_3}{2} + \omega\hat{a}^{\dagger}\hat{a} + \frac{\lambda^2}{\Delta}\Bigl(\hat{\sigma}_3\hat{a}^{\dagger}\hat{a} + \ket{e}\bra{e}\Bigr).
\end{equation}
In the next section we shall make use of these results to develop a framework for the time evolution of the atom-field system. 

\section{Time Evolution}
\subsection{Background}
The most general (pure) initial states of the atom and the field are given by linear superpositions of all the accessible states
\begin{eqnarray}
\ket{\psi(0)}_a & = & C_g\ket{g} + C_e\ket{e}  \label{eq:GeneralAtom} \\
\ket{\psi(0)}_f & = & \sum_{n=0}^{\infty}C_n\ket{n}  \label{eq:GeneralField},
\end{eqnarray}
where the $C_x$ are arbitrary coefficients, satisfying the normalisation conditions $|C_g|^2 +|C_e|^2 = 1$ and $\sum_{n=0}^{\infty}|C_n|^2 = 1$. The total initial state of the bipartite atom-field system is then given by
\begin{equation} \label{eq:State}
\ket{\psi(0)} = \ket{\psi(0)}_a \otimes \ket{\psi(0)}_f.
\end{equation}\\

\subsubsection{Resonant Interaction, $\Delta = 0$}
Allowing the $C$ coefficients to vary with time, we directly solved the interaction picture Schr\"odinger equation
\begin{equation} \label{eq:TDSE}
\frac{d\ket{\psi(t)}}{dt} =  \hat{H}_{I}\ket{\psi(t)}
\end{equation}
for the case of no detuning ($\Delta = 0$), where the interaction Hamiltonian takes the simple form 
\begin{equation} \label{eq:NoDet}
\hat{H}_{I} = \lambda(\hat{\sigma}_+\hat{a} + \hat{\sigma}_-\hat{a}^{\dagger})
\end{equation}
as can be directly seen by substituting $\Delta = 0$ into \eqref{eq:JCM_int}. We found the analytical solution to \eqref{eq:TDSE} at an arbitrary time $t$ as
\begin{widetext}
\begin{equation} \label{eq:GeneralTDSE}
\begin{split}
\ket{\psi(t)} =  \sum_{n=0}^{\infty} \Bigl[  \Bigl( C_eC_n\cos(\lambda t \sqrt{n+1}) - i C_gC_{n+1}\sin(\lambda t \sqrt{n+1})\Bigr)&\ket{e} \\
+  \Bigl(- i C_eC_{n-1}\sin(\lambda t \sqrt{n}) + C_gC_n\cos(\lambda t \sqrt{n})\Bigr)&\ket{g} \Bigr] \otimes \ket{n}.
\end{split}
\end{equation}
\end{widetext}
An equivalent but more versatile and powerful approach (especially when working with density matrices, see below) is to explicitly calculate the time-evolution operator
\begin{equation} \label{eq:EvoDef}
\hat{U} = e^{-i\hat{H}_{I}t}.
\end{equation}
By expanding the exponential into its power series and calculating the matrix elements, again using the Hamiltonian given in \eqref{eq:NoDet}, we showed that the full evolution operator of the JCM at resonance, written in the atomic basis $\{\ket{e},\ket{g}\}$, is given by \cite{Isham1995,Barnett2003}
\begin{equation} \label{eq:Evo}
\hat{U}(t) = 
\begin{pmatrix}
\cos(\lambda t \sqrt{\hat{a}\hat{a}^{\dagger}}) &
-i \hat{a}\frac{\sin(\lambda t \sqrt{\hat{a}^{\dagger}\hat{a}})}{\sqrt{\hat{a}^{\dagger}\hat{a}}} \\
-i \hat{a}^{\dagger}\frac{\sin(\lambda t \sqrt{\hat{a}\hat{a}^{\dagger}})}{\sqrt{\hat{a}\hat{a}^{\dagger}}} &
\cos(\lambda t \sqrt{\hat{a}^{\dagger}\hat{a}})
\end{pmatrix}.
\end{equation}
Applying \eqref{eq:Evo} to the initial state \eqref{eq:State}
\begin{equation} \label{eq:Evo2}
\ket{\psi(t)} = \hat{U}(t)\ket{\psi(0)}
\end{equation}
exactly reproduces the result obtained in \eqref{eq:GeneralTDSE}, as expected.\\
Considering equation \eqref{eq:GeneralTDSE}, it is straightforward to analyse the types of interaction that are taking place in the resonant case. The first line of \eqref{eq:GeneralTDSE} shows that the excited atomic state $\ket{e}$ can interact with the field number state $\ket{n}$ by exciting the field into the higher number state $\ket{n+1}$ while the atom itself drops into its ground state $\ket{g}$ to conserve the total excitation number. Alternatively this process can take place in reverse, whereby the atom gets excited $\ket{g}\rightarrow\ket{e}$ by absorbing a photon from the field $\ket{n}\rightarrow\ket{n-1}$.

\subsubsection{Off-Resonant Interaction, $\Delta \gg \lambda$}
The off-resonant interaction behaves distinctly different in comparison to this. Starting again with the definition in \eqref{eq:EvoDef} and using the off-resonant interaction Hamiltonian \eqref{eq:LargeDet_eff} it is straightforward to find the corresponding evolution operator. The general initial state \eqref{eq:State} is an eigenstate of this $\hat{H}_I$. Hence the exponential involving $\hat{H}_I$ that appears in \eqref{eq:Evo2} can be replaced by the equivalent exponential with the operator replaced by its eigenvalues. This gives the state of the system in the off-resonant case at time $t$ as
\begin{equation} \label{eq:GeneralTDSEOff}
\begin{split}
\ket{\psi(t)} =  \sum_{n=0}^{\infty} C_n \Bigl[ C_e e^{-i\chi(n+1)t}\ket{e} +  C_g e^{i\chi n t}\ket{g} \Bigr]  \ket{n}, 
\end{split}
\end{equation}
where we have defined
\begin{equation} \label{eq:chi}
\begin{split}
\chi = \frac{\lambda^2}{\Delta} 
\end{split}
\end{equation}
for convenience.\\
It is immediately obvious that \eqref{eq:GeneralTDSEOff} represents a very different interaction compared to the resonant case \eqref{eq:GeneralTDSE}. Here, no energy or excitation exchange between the atom or the field take place. Only the relative phases are affected and rotate in time.\\
In the following subsection we will use the results \eqref{eq:GeneralTDSE} and \eqref{eq:GeneralTDSEOff} to look at some illustrative exemplary systems.\\

\subsection{Two Simple Examples}
Having derived the general time evolution of an arbitrary system in the two distinct cases, we can now focus on some particular examples and analyse their properties. The coherent state $\ket{\alpha}$ for which the $C_n$ coefficients take the form
\begin{equation} \label{eq:coherent}
C_n = e^{-\frac{|\alpha|^2}{2}}\frac{\alpha^n}{\sqrt{n!}},
\end{equation}
where $\alpha$ is related to the fields mean photon number $\bar{n}$ by $\bar{n} = |\alpha|^2$, is of particular interest. These states are in some sense the ``most classical'' field states, since they obey the classical equations of motion of the simple harmonic oscillator and they can also serve as an idealisation of the light emitted by a laser \cite{Gerry2004}.\\
It will also prove useful later on to define the notion of the density operator $\hat{\rho}(t)$, which for a pure state $\ket{\psi(t)}$ is simply given by the outer product of the state with itself
\begin{equation} \label{eq:density}
\hat{\rho}(t) = \ket{\psi(t)}\bra{\psi(t)}.
\end{equation}
In terms of the density operator, the expectation value of an observable $A_i$ of subsystem $i$ (here either atom or field) is given by
\begin{equation} \label{eq:expect}
\braket{A_i} = Tr[\hat{\rho}_i\hat{A}_i],
\end{equation}
where $\hat{\rho}_i$ is the density operator of the subsystem, given by the partial trace of $\hat{\rho}$ over the other subsystem \cite{Isham1995}.\\
Equipped with these tools we can now look at an atom, initially in the excited state ($C_e=1$,  $C_g=0$), that interacts with a coherent field of the general form \eqref{eq:GeneralField}. Substituting the atomic coefficients into \eqref{eq:GeneralTDSE} and taking the outer product gives the density matrix $\hat{\rho}$ of the system. To find the atomic density matrix $\hat{\rho}_a$ we take the partial trace over the field
\begin{equation} \label{eq:AtomDensDef}
\hat{\rho}_a = \sum_{n=0}^{\infty}\braket{n | \hat{\rho} | n}.
\end{equation}
\begin{figure}[t]
\begin{center}
\includegraphics[width = \columnwidth]{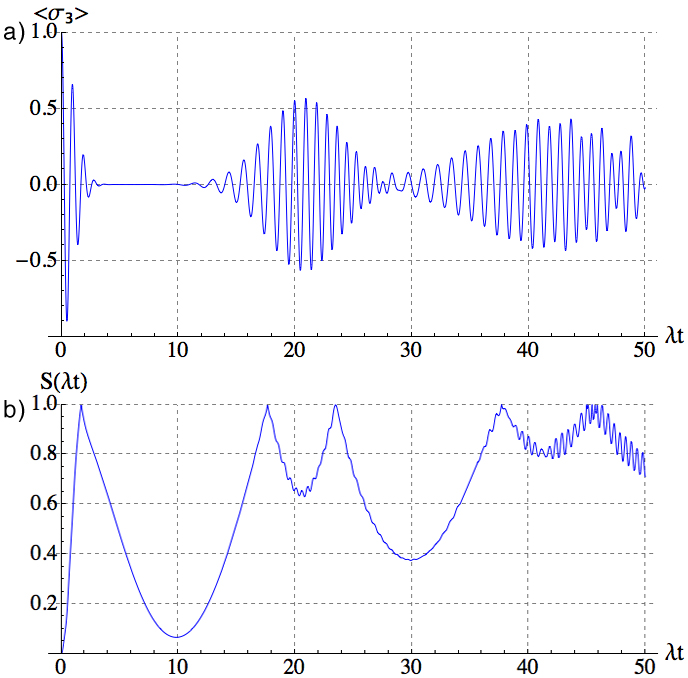}
\caption{a) Atomic inversion of an initially excited atom interacting with a coherent field with $\bar{n}=10$, plotted against the scaled interaction time $\lambda t$. b) Von Neumann entropy of the system. Series in this and the following figures were evaluated up to $C_{100}$ unless otherwise stated.}
\label{Inversion}
\end{center}
\end{figure}
We can use this result and \eqref{eq:expect} to calculate the atomic inversion 
\begin{equation} \label{eq:Inversion}
\begin{split}
\braket{\sigma_3} &= Tr[\hat{\rho}_a\hat{\sigma}_3] \\
&=  \sum_{n=0}^{\infty}|C_n|^2\cos(2\lambda t \sqrt{n+1})
\end{split}
\end{equation} 
which describes the state of the atom at any given time. $\braket{\sigma_3} = 1$ corresponds to an atom that is fully excited, whereas $\braket{\sigma_3} = -1$ describes an atom in the ground state. Intermediate values imply superpositions of both states. Figure \ref{Inversion}a shows the inversion against scaled time $\lambda t$ for an excited atom that interacts with a coherent field with average photon number $\bar{n} = 10$. The figure clearly shows the collapses and revivals that are characteristic for the JCM. These are often considerered as a proof for the quantised nature of radiation \cite{Shore1993}.\\
Directly related to the atomic inversion is the expected photon number in the field $\braket{n}$, where $\hat{n}=\hat{a}^{\dagger}\hat{a}$ is the number operator. We expect the total excitation number $N = \frac{\braket{\sigma_3}+1}{2}+\braket{n}$ to be constant to conserve energy and match the nature of the interaction described above. Direct calculation confirms this expectation, giving a graph for $\braket{n}$ that is essentially a shifted atomic inversion, mirrored along the $\braket{\sigma_3}=0$ axis. \\
Another interesting quantity is the von Neumann entropy of subsystem $i$
\begin{equation} \label{eq:Entropy}
S_i = -Tr[\hat{\rho}_i\log_2{\hat{\rho}_i}]
\end{equation}
which serves an an entanglement measure for pure states \cite{Eisert1998,Plenio2005a,Vedral1998,Plenio1998,Hessian2009}, giving information about the amount of quantum correlations between the atom and the field. The von Neumann entropy is less straight forward to calculate since it involves the logarithm of the density operator. To compute this we have to diagonalise $\hat{\rho}$, which can be done using Schmidt decomposition \cite{Gerry2004}. The result is shown in Figure \ref{Inversion}b. One can see that the system starts with no entanglement ($S=0$) as is expected from a pure initial state \cite{Phoenix1991c}, but rapidly evolves into an almost fully entangled state ($S\approx1$) and then oscillates at an intermediate entropy. Hence, by tuning the interaction time of the atom with the field to match exactly the entropy maximum, a strongly entangled atom-field system could be created, which could then be used as a qubit-pair for purposes in quantum information processing (see section V). However, the entropy peaks are very narrow and it would prove experimentally difficult to achieve the correct interaction time. Note that although we have considered the atomic subsystem here, the result is generally true for the total system, since it starts out in a pure state. Hence the subsystems have equal entropy throughout their subsequent evolution \cite{Gerry2004,Boukobza2005}.\\
For the case of large detuning we found the initial state
\begin{equation} \label{eq:StateOff}
\ket{\psi(0)} = \frac{1}{\sqrt{2}}\bigl(\ket{g}+e^{i\phi}\ket{e}\bigr)\otimes\ket{\alpha}
\end{equation}\\
with $C_g = \frac{1}{\sqrt{2}}$, $C_e = \frac{1}{\sqrt{2}}e^{i\phi}$ (for some arbitrary phase $\phi$) and the atom in the coherent state to have particularly interesting properties. Substituting the coefficients into \eqref{eq:GeneralTDSEOff} gives the state, and hence the full density operator, at time $t$. Tracing over the field, we find the atomic density operator 
\begin{equation} \label{eq:AtomDens}
\begin{split}
\hat{\rho}_a = &\frac{1}{2}\Bigl[ \ket{g}\bra{g} + \ket{e}\bra{e} \\
& + \ket{g}\bra{e}\sum_{n=0}^{\infty}|C_n|^2e^{-i(\chi t(2n-1) + \phi)} \\
& + \ket{e}\bra{g}\sum_{n=0}^{\infty}|C_n|^2e^{+i(\chi t(2n-1) + \phi)} \Bigr].
\end{split}
\end{equation}
It is easy to see from this that the atomic inversion is constant, and in fact zero, as is to be expected for the off-resonant interaction, since we showed above that no excitation exchange takes place between atom and field. \\
A much more interesting quantity is the entropy. Equation \eqref{eq:AtomDens} can again be diagonalised by means of Schmidt decomposition. The explicit form for the entropy is then given by
\begin{equation}
S(t) = -|g_+|^2\log_2|g_+|^2 - |g_-|^2\log_2|g_-|^2 \label{eq:Schmidt} 
\end{equation}
where
\begin{eqnarray}
g_{\pm} & = & \frac{1}{2}\Bigl[1 \pm \sqrt{x_1^2 + x_2^2 + x_3^2}\Bigr] \label{eq:Schmidt2}, \\
x_1 & = & \sum_{n=0}^{\infty}|C_n|^2\cos[\chi t(1-2n) - \phi)], \\
x_2 & = & \sum_{n=0}^{\infty}|C_n|^2\sin[\chi t(1-2n) - \phi)], \\
x_3 & = & 0.
\end{eqnarray}
\begin{figure}[b]
\begin{center}
\includegraphics[width = \columnwidth]{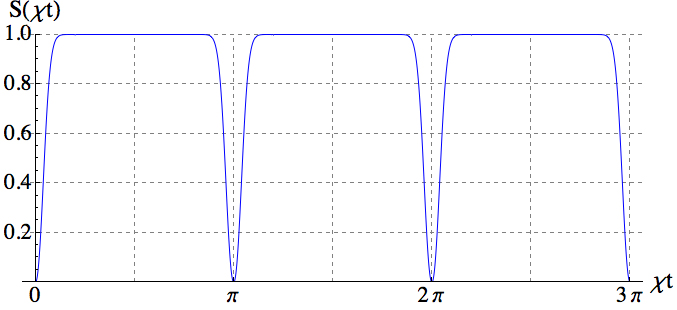}
\caption{Von Neumann entropy $S$ of an atom in the state $\frac{1}{\sqrt{2}}\bigl(\ket{g}+e^{i\phi}\ket{e}\bigr)$ interacting with a coherent field with $\bar{n}=10$.}
\label{Entropy}
\end{center}
\end{figure}
Figure \ref{Entropy} shows the entropy plotted against scaled time $\chi t$ for a coherent state with $\bar{n} = 10$. Interestingly the initially unentangled system rapidly reaches a near maximally entangled state in which it remains except for interaction times close to $\chi t = n\pi$. This result suggests that this particular interaction could be used as a reasonably robust method for generating entanglement between the atom and the field, since it is remarkably insensitive to slight variations in the interaction time in the proximity of $\chi t \approx (n+\frac{1}{2})\pi$. Numerical analysis indicates that the result is independent of the phase $\phi$.

\section{Entangling two Cavity Modes}
In the previous section we described two interactions that lead to entanglement between an atom and an electromagnetic field mode. In this section we shall look at a different method that allows for the very precise generation of maximally entangled states between two optical cavities.
\begin{figure}[t]
\begin{center}
\includegraphics[width = \columnwidth]{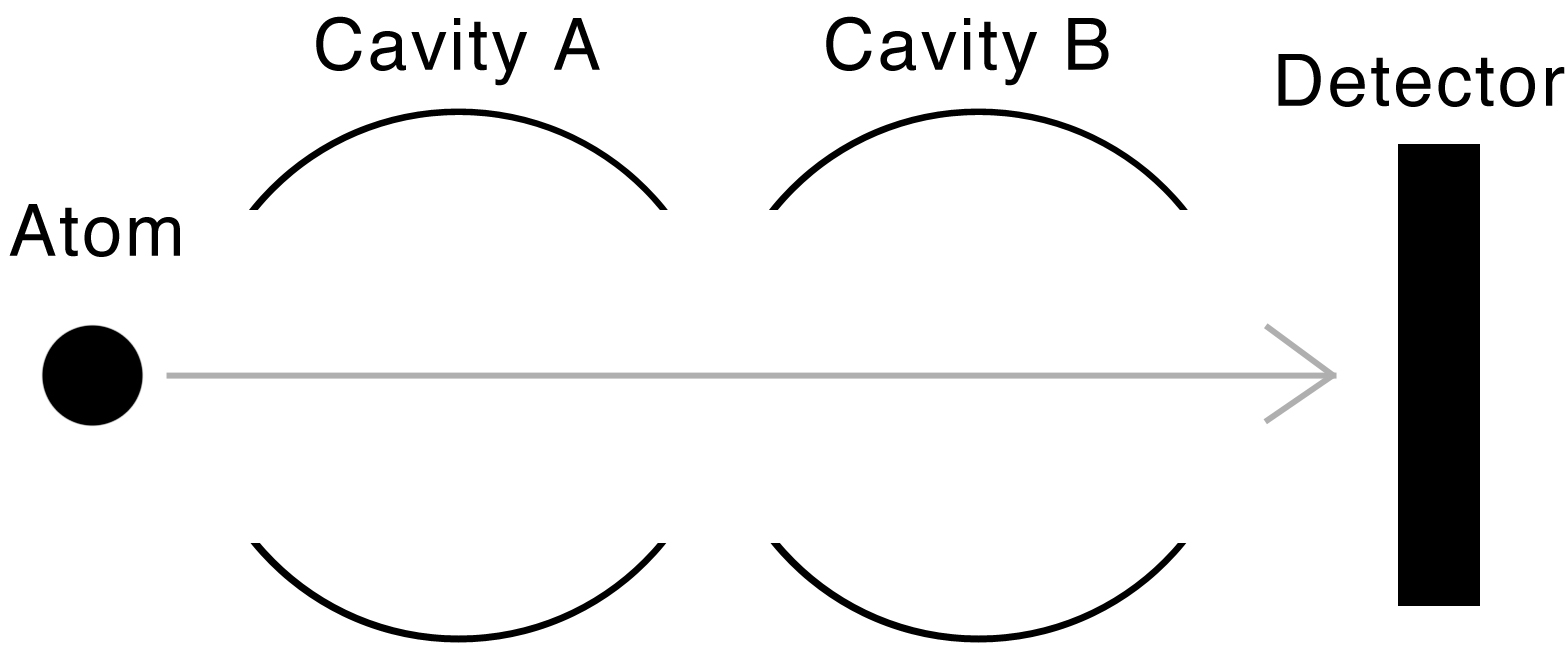}
\caption{Basic experimental setup used for the two schemes. An initially prepared atom passes through two consecutive optical cavities and is then measured via a detector. }
\label{Layout}
\end{center}
\end{figure}
\subsection{Basic Idea}
The basic concept was initially developed by Browne and Plenio \citep{Browne2002a}. The underlying experimental setup is shown in Figure \ref{Layout}. Initially an atom is prepared in a certain state. It is then sent through two consecutive cavities A and B and allowed to interact with them for a certain time that is determined by the time it takes the atom to traverse the cavities. The cavities themselves do not directly interact with each other. Once the atom has passed through both cavities it hits a detector which makes measurements in the $\{\ket{e}, \ket{g}\}$ basis. This measurement collapses the state of the two cavities into a new state, which under the correct initial conditions and interaction times, is a maximally entangled Bell state.

\subsection{Generating Bell States}
The four Bell states, given by
\begin{eqnarray}
\ket{\Psi^\pm}_{AB} & = & \frac{1}{\sqrt{2}}(\ket{0}_A\ket{1}_B \pm \ket{1}_A\ket{0}_B) \label{eq:Bell1} \\
\ket{\Phi^\pm}_{AB} & = & \frac{1}{\sqrt{2}}(\ket{0}_A\ket{0}_B \pm \ket{1}_A\ket{1}_B)\label{eq:Bell2}
\end{eqnarray}
are the maximally entangled states of a bipartite system \cite{Vedral1998a,Plenio1998,Browne2002a,Preskill1998,Isham1995}. They are of fundamental importance since they form the basis of many protocols in quantum information processing, such as quantum teleportation which we will discuss in section V. Hence it is crucial that we are able to efficiently produce these states in the laboratory in order to realise these protocols.\\
\subsubsection{Browne-Plenio Scheme}
The scheme developed by Browne and Plenio makes it possible to generate one specific Bell state, namely $\ket{\Psi^+}$. In this scheme the atom starts in the excited state $\ket{e}$ and both cavities, with frequencies equal to the energy splitting of the atom, i.e. $\Delta = 0$, are initially in the vacuum state $\ket{0}$, giving a combined initial state of 
\begin{equation}\label{eq:PlenioState} 
\ket{\psi_{init}} =  \ket{e}\otimes\ket{0}_A\otimes\ket{0}_B.
\end{equation}
Since the atom traverses one cavity at a time, first A and then B, we can make use of the general solution to the time-dependent Schr\"odinger equation we derived earlier in equation \eqref{eq:GeneralTDSE}. First, we can completely ignore cavity B and only think about the state of the system right after the atom leaves cavity A, since we know B does not take part in this interaction and hence just remains in the state $\ket{0}_B$. Therefore, substituting $C_e=1$, $C_g=0$, $C_0 = 1$ and $C_n = 0~\forall~n > 0$ into \eqref{eq:GeneralTDSE} we find that the total state of our system is
\begin{equation}\label{eq:PlenioStateInter} 
\ket{\psi_{inter}} = \Bigl[ \cos(\lambda\tau)\ket{e}\ket{0}_A - i\sin(\lambda\tau)\ket{g}\ket{1}_A\Bigr]\ket{0}_B
\end{equation}
after it has interacted with cavity A for a time $t=\tau$.\\
Now the atom will enter cavity B. Since it now only interacts with B and not A, we can redefine our initial conditions for this second interaction as 
\begin{eqnarray}
C'_e & = & \cos(\lambda\tau)\ket{0}_A \\
C'_g & = & - i\sin(\lambda\tau)\ket{1}_A \\
C'_n & = & \begin{cases} 1 & \text{if } n = 0 \\
   0       & \text{if } n > 0 \end{cases}
\end{eqnarray}
and use the same method of substitution into \eqref{eq:GeneralTDSE} to arrive at the solution
\begin{equation}\label{eq:PlenioFinal} 
\begin{split}
\ket{\psi_{final}}  =   \Bigl[ \cos^2(\lambda\tau)\ket{0}_A\ket{0}_B\Bigr]  & \ket{e} \\
- i\sin(\lambda\tau)  \Bigl[\cos(\lambda\tau)\ket{0}_A\ket{1}_B + \ket{1}_A\ket{0}_B \Bigr] & \ket{g},
\end{split}
\end{equation}
assuming the interaction time is again $t=\tau$.\\
As described above, we now carry out a measurement of the atom in the $\{\ket{e}, \ket{g}\}$ basis. There are two possible outcomes. If we measure the atom in the excited state, the cavities collapse back into their initial state $\ket{\psi}_{cav}~=~\ket{0}_A\ket{0}_B$, which is not a useful state in itself, but it allows the scheme to be restarted immediately. The interesting case is if we measure the atom in its ground state. The the two cavity system then collapses into the (unnormalised) state
\begin{equation} \label{eq:cav}
\ket{\psi_{cav}}=  \cos(\lambda\tau)\ket{0}_A\ket{1}_B + \ket{1}_A\ket{0}_B.
\end{equation}
\begin{figure}[t]
\begin{center}
\includegraphics[width = \columnwidth]{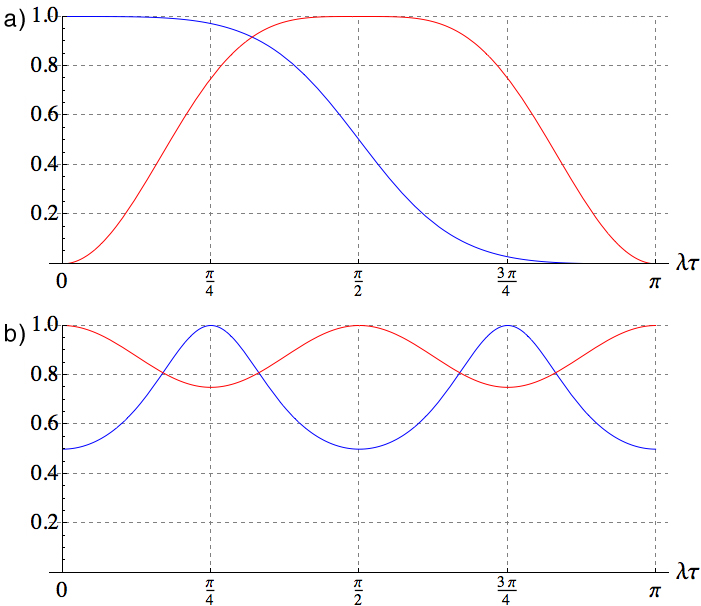}
\caption{Fidelity $F$ (blue) and success probability $P$ (red) plotted against scaled interaction time $\lambda\tau$ for a) the Browne-Plenio scheme and b) the scheme we propose. }
\label{Fidelity}
\end{center}
\end{figure}
By comparing this result to the Bell states in equation \eqref{eq:Bell1} it is obvious that this state approaches the state $\ket{\Psi^+}$ for interaction times $\lambda\tau\rightarrow0$. To further analyse how close the two-cavity-state is to $\ket{\Psi^+}$ we can calculate the fidelity $F=|\braket{\psi_{cav}|\Psi^+}|^2$, where we first have to normalise our state $\ket{\psi_{cav}}$.
\begin{equation} \label{eq:Fidelity}
F_{BP} = \frac{1}{2} + \frac{\cos(\lambda\tau)}{\cos^2(\lambda\tau) + 1}
\end{equation}
The subscript $BP$ labels this as the fidelity for the Browne-Plenio scheme. We see that as expected $F_{BP}\rightarrow1$ as $\lambda\tau\rightarrow0$.\\
However, we also have to consider how likely our experiment is to succeed, that is, with what probability we will measure the atom to be in the ground state. This probability is
\begin{equation} \label{eq:Probab}
\begin{split}
P_{BP} &= 1- |\braket{e|\psi_{final}}|^2 \\
&=  1 - \cos^4(\lambda\tau).
\end{split}
\end{equation}
Figure \ref{Fidelity}a shows the fidelity and success rate against scaled interaction time $\lambda\tau$. The virtue of this scheme is its robustness. The fidelity curve is very flat for small $\lambda\tau$, and even at $\lambda\tau$ as high as $\frac{\pi}{4}$, where $P_{BP}=0.75$, we still have $F_{BP}\approx0.97$ which could be high enough for many applications. However, it can be seen that we always have to compromise when using this scheme. The fidelity does go to $1$ as $\lambda\tau\rightarrow0$, but in this limit the success probability goes to zero. This means that we can get arbitrarily close to a Bell state by sacrificing success probability, but a perfect Bell state is theoretically impossible to achieve. 

\subsubsection{New Scheme}
We propose a new scheme, which allows us to generate Bell states with arbitrary fidelity and non-vanishing success probability. Initially, we showed that using the same approach of solving the Schr\"odinger equation in two separate steps, the cavity state after interaction of an arbitrary atom with arbitrary cavity fields of the form of equations \eqref{eq:GeneralAtom} and \eqref{eq:GeneralField} (the latter of which could be prepared as described by Law and Eberly \cite{Law1996}), and after the atom has been measured in the ground state, is given by the (unnormalised) state
\begin{widetext}
\begin{equation} \label{eq:General2Cav}
\begin{split}
\ket{\psi_{cav}} =  \sum_{n=0}^{\infty}\sum_{m=0}^{\infty}   \Bigl[ &
i C_eC_n^AC_{m-1}^B\cos(\Omega_{n+1}^A)\sin(\Omega_{m}^B)
 + C_gC_{n+1}^AC_{m-1}^B\sin(\Omega_{n+1}^A)\sin(\Omega_{m}^B) \\
& +i C_eC_{n-1}^AC_{m}^B\sin(\Omega_{n}^A)\cos(\Omega_{m}^B) 
 - C_gC_{n}^AC_{m}^B\cos(\Omega_{n}^A)\cos(\Omega_{m}^B)
 \Bigr] \ket{n}_A \otimes \ket{m}_B,
\end{split}
\end{equation}
\end{widetext}
where we have defined
\begin{equation} \label{eq:omega}
\Omega_n^A = \lambda\tau_A\sqrt{n}
\end{equation}
and an equivalent expression for $\Omega_m^B$. $\tau_A$ and $\tau_B$ are the interaction times with cavity A and cavity B respectively. It is straightforward to show that on substitution of the correct coefficients and by setting $\tau_A = \tau_B = \tau$ the state \eqref{eq:General2Cav} reduces to the result obtained in \eqref{eq:cav}, as expected.\\
The new scheme we propose starts with an equally simple state as the scheme described above. We suggest to start with an atom in the ground state ($C_g=1$, $C_e=0$), the first cavity in the number state $\ket{1}_A$ and the second cavity again in the vacuum state $\ket{0}_B$. This cavity state can be prepared by starting from the one described above, with both cavities in the vacuum state as follows. We have derived in equation \eqref{eq:PlenioStateInter} the state of the system after cavity A has interacted with an excited atom. If we now measure the atom and find it in its ground state, we have prepared the cavities in the desired state $\ket{1}_A\ket{0}_B$. If we measure $\ket{e}$ instead, the cavity reverts back to the original state $\ket{0}_A$ and we can try again using another atom. Once we have successfully prepared the cavities using these excited ``auxiliary atoms'' we can begin with the actual scheme, using a ground state atom.\\ 
The motivation behind the initial state $\ket{g}\ket{1}_A\ket{0}_B$ is the following. According to the interaction types described in section III, in cavity A the ground state atom can either pass through without any energy and hence excitation exchange with the field, or it can absorb the cavity photon, which would bring it into its excited state. This leaves the atom in a superposition of excited and ground state and the cavity in a superposition of $\ket{0}_A$ and $\ket{1}_A$. Now, as the atom enters cavity B, its excited component can interact with the field by a similar, but in some sense reverse process compared to cavity A, namely deexciting via the emission of a photon. This then leaves cavity B also in a superposition of $\ket{0}_B$ and $\ket{1}_B$, and in fact the combined cavity state can again be turned into a Bell state by precise tuning of the interaction time.\\
Upon substitution of the coefficients into \eqref{eq:General2Cav} we find that the cavity will be in the (unnormalised) state
\begin{equation} \label{eq:cav2}
\ket{\psi_{cav}}=  \sin^2(\lambda\tau)\ket{0}_A\ket{1}_B - \cos^2(\lambda\tau)\ket{1}_A\ket{0}_B
\end{equation}
after we measure the atom in its ground state. We have again assumed equal interaction times $\tau$ in both cavities. We can compare this to the Bell states in \eqref{eq:Bell1} to find that $\ket{\psi_{cav}}=\ket{\Psi^-}$ when $\sin^2(\lambda\tau) = \cos^2(\lambda\tau)$. This has the solution $\lambda\tau = \frac{\pi}{4}(2n-1)$. The shortest interaction time that gives the desired state is thus $\lambda\tau = \frac{\pi}{4}$.\\
Normalising $\ket{\psi_{cav}}$ we can again compute the fidelity of our state in comparison with the Bell state, for an arbitrary interaction time.
\begin{equation} \label{eq:Fidelity2}
F = \frac{1}{2} + \frac{\sin^2(2\lambda\tau)}{\cos(4\lambda\tau) + 3}
\end{equation}
Comparing this to \eqref{eq:Fidelity} we immediately see one obvious advantage of our scheme, namely that the fidelity never falls below 0.5. The success probability is given by 
\begin{equation} \label{eq:Probab2}
P = 1 - \frac{1}{4}\sin^2(2\lambda\tau).
\end{equation}
Both $F$ and $P$ are plotted in Figure \ref{Fidelity}b. We see that although the fidelity maximum and success probability minimum coincide, as was the case in the Browne-Plenio scheme, our scheme allows the creation of states that are arbitrarily close to the desired state. The only limiting factor here is the experimental imprecision in tuning the interaction time perfectly to its ideal value. But assuming that this is a problem that can be overcome experimentally, we have a scheme that can generate Bell states with a success probability of $P=0.75$, as opposed to the vanishing probability of the original scheme. The disadvantage of our scheme is however the stronger curvature of the fidelity curve near its maximum, which implies that it is not as robust against variations in the interaction time. This is however only an experimental limitation.

\subsection{Interaction Times}
\begin{figure}[b]
\begin{center}
\includegraphics[width = 0.7\columnwidth]{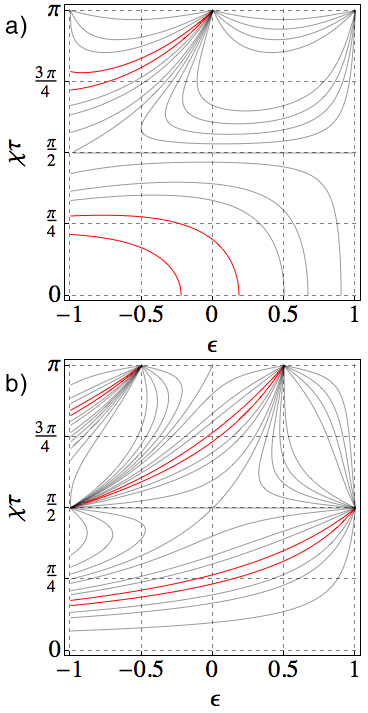}
\caption{Contour plots of the fidelity as a function of interaction time $\chi t$ and deviation $\epsilon$ for a) the Browne- Plenio scheme and b) the scheme we propose. The red contour corresponds to a fidelity of 0.99. The remaining fidelity contours are at 0.1,~0.2,~...,~0.9.}
\label{FidelityNew}
\end{center}
\end{figure}

Browne and Plenio also considered the effect of different interaction times in the two cavities due to experimental imprecision \cite{Browne2002a}. This could be caused for example by cavities of slightly different dimensions or by the atom being slightly deflected while traversing the cavities, hence not taking exactly the same straight path in both cavities. Browne and Plenio proceed by assuming that the interaction time in cavity A is $\tau_A = \tau$ and the time in cavity B is $\tau_B = \tau(1-\epsilon)$, where we have so far not made any explicit assumption about the size of $\epsilon$ other than $\epsilon < 1$.\\
Utilising again our general solution \eqref{eq:General2Cav} we can immediately incorporate the now different interaction times into the previous solution for the Browne-Plenio scheme. Thus, finding the new cavity state after the atom is measured in its ground state, we now arrive at a new fidelity
\begin{equation} \label{eq:Fidelity3}
F'_{BP} = \frac{1}{2} + \frac{\cos(\lambda\tau)\sin(\lambda\tau)\sin(\lambda\tau(1-\epsilon))}{\cos^2(\lambda\tau)\sin^2(\lambda\tau(1-\epsilon)) + \sin^2(\lambda\tau)}.
\end{equation}
A similar treatment of our scheme gives the fidelity
\begin{equation} \label{eq:Fidelity3}
F' = \frac{1}{2} + \frac{\sin(2\lambda\tau)\sin(2\lambda\tau(1-\epsilon))}{2\cos(2\lambda\tau)\cos(2\lambda\tau(1-\epsilon))+2}.
\end{equation}
The two fidelities are plotted as functions of $\lambda\tau$ and $\epsilon$ in Figures \ref{FidelityNew}a and \ref{FidelityNew}b respectively.
We can see that both schemes exhibit extended areas for which $F>0.99$, which implies a certain robustness against experimental inaccuracies. In line with Figure \ref{Fidelity}, we see that the Browne-Plenio scheme has a flatter fidelity function. The contour plots also demonstrate the interesting possibility to use different interaction times for cavity A and B to achieve the desired states. In fact our scheme allows us to find a suitable $\epsilon$ for any interaction time $\frac{\pi}{6}\lesssim\tau_A\leq\pi$.

\subsection{Thermal States}
We can further extend both schemes by asking what would happen if we allowed the system to thermally interact with its environment. This can be done by allowing either the atom or the cavities to be in the thermal state, which is described by the density operator
\begin{equation} \label{eq:Thermal}
\hat{\rho}_{Th} =  \frac{e^{-\beta\hat{H}}}{Tr[e^{-\beta\hat{H}}]},
\end{equation}
where $\beta= (k_BT)^{-1}$ with $k_B$ being the Boltzmann constant, $T$ the temperature, and $\hat{H}$ the atomic or field Hamiltonian discussed in section II \cite{Isham1995}. For the atom it can be shown \cite{Gerry2004,Boukobza2005} that this is equal to 
\begin{equation} \label{eq:ThermalAtom}
\hat{\rho}_{Th}^A =  P_e\ket{e}\bra{e} + P_g\ket{g}\bra{g},
\end{equation}
with 
\begin{eqnarray}
P_e & = & (e^{\beta\omega_0}+1)^{-1} \\
P_g & = & (e^{-\beta\omega_0}+1)^{-1}.
\end{eqnarray}
A similar result can be obtained for the thermal states of the field, where
\begin{equation} \label{eq:ThermalField}
\hat{\rho}_{Th}^F =  \sum_{n=0}^{\infty}P_n\ket{n}\bra{n}
\end{equation}
with
\begin{equation} \label{eq:ThermalFieldCoeff}
P_n =  \frac{\bar{n}^n}{(1+\bar{n})^{n+1}}
\end{equation}
and
\begin{equation} \label{eq:nbar}
\bar{n} = (e^{\beta\omega}-1)^{-1}.
\end{equation}
Both these thermal states are distinctly different from the states we have considered so far. They are intrinsically mixed states, representing statistical mixtures of all the accessible states. Hence they can not be separated into pure states as was the case with the initial states considered so far. The only way to describe the thermal states is via the density operators introduced here.\\
This means we can now no longer rely on the solution of the Schr\"odingier equation \eqref{eq:GeneralTDSE}, but have to make use of the evolution operator \eqref{eq:Evo}. The density operator then evolves according to \cite{Isham1995}
\begin{equation} \label{eq:DensityEvo}
\hat{\rho}(t) = \hat{U}(t)\hat{\rho}(0)\hat{U}^{\dagger}(t).
\end{equation}
\subsubsection{Browne-Plenio Scheme}
First we look at the Browne-Plenio scheme where we now allow the two cavity modes to be in thermal states. The atom still starts out in the pure state $\ket{e}$ which can be described by the density operator $\hat{\rho}_a = \ket{e}\bra{e}$. We can again treat the two interactions individually. The density operator for the joint system of the atom and cavity A is  
\begin{equation} \label{eq:DensityInit}
\begin{split}
\hat{\rho}(0) & = \hat{\rho}_a \otimes \hat{\rho}_{Th_A}^F \\
& = \begin{pmatrix}
\hat{\rho}_{Th_A}^F & 0 \\
0 & 0
\end{pmatrix},
\end{split}
\end{equation}
where the second line refers to the atomic basis. We can now use equation \eqref{eq:DensityEvo} to find the density operator of the system after the atom has interacted with the thermal field in cavity A for $t=\tau$. The resulting $4\times4$-matrix can be defined as
\begin{equation} \label{eq:DensityInit}
\hat{\rho}(\tau)  = \begin{pmatrix}
\hat{\rho}_{ee} & \hat{\rho}_{eg} \\
\hat{\rho}_{ge} & \hat{\rho}_{gg}
\end{pmatrix}.
\end{equation}
The total density matrix before the interaction with cavity B is then $\hat{\rho}(\tau)\otimes \hat{\rho}_{Th_B}^F$, i.e.
\begin{equation} \label{eq:DensityInter}
\hat{\rho}'(0)  = \begin{pmatrix}
\hat{\rho}_{ee}\otimes\hat{\rho}_{Th_B}^F & \hat{\rho}_{eg}\otimes\hat{\rho}_{Th_B}^F \\
\hat{\rho}_{ge}\otimes\hat{\rho}_{Th_B}^F & \hat{\rho}_{gg}\otimes\hat{\rho}_{Th_B}^F
\end{pmatrix}
\end{equation}
in the atomic basis. Using the evolution operator to once more evolve this state by a time $\tau$ gives the final state of the full system, $\hat{\rho}_{final}$. When we measure the atom in the ground state, the remaining cavity density matrix is $\hat{\rho}_{cav} = \braket{g | \hat{\rho}_{final} | g}$. The result can be written as
\begin{equation} \label{eq:DensityCav}
\begin{split}
\hat{\rho}_{cav}  =   (&\hat{C}\hat{\rho}_{Th}^F\hat{C})_A\otimes(\hat{S}\hat{\rho}_{Th}^F\hat{S}')_B \\
 + (&\hat{S}\hat{\rho}_{Th}^F\hat{C})_A\otimes(\hat{C}'\hat{\rho}_{Th}^F\hat{S}')_B \\
 + (&\hat{C}\hat{\rho}_{Th}^F\hat{S}')_A\otimes(\hat{S}\hat{\rho}_{Th}^F\hat{C}')_B \\
 + (&\hat{S}\hat{\rho}_{Th}^F\hat{S}')_A\otimes(\hat{C}'\hat{\rho}_{Th}^F\hat{C}')_B
\end{split}
\end{equation}
where $\hat{C}$, $\hat{S}$, $\hat{C}'$ and $\hat{S}'$ are the matrix elements of the evolution operator $\hat{U}  = \Bigl(\begin{smallmatrix}
\hat{C} & \hat{S}' \\
\hat{S} &  \hat{C}'
\end{smallmatrix}\Bigr)$ (c.f. equation \eqref{eq:Evo}). The exact evaluation of $\hat{\rho}_{cav}$ following the procedure above is straightforward albeit very laborious and is left out here.\\
After truncating the series describing the thermal states for $n > 1$ (since then the $P_n$ are of negligible size under normal conditions, see below), $\hat{\rho}_{cav}$ can be expressed as a $9\times9$-matrix in the field basis $\{\ket{0}_A\ket{0}_B, ... , \ket{2}_A\ket{2}_B\}$.\\
At this point we are no longer able to simply compare this state with the Bell state to judge how much entanglement we have. Furthermore, the von Neumann entropy is not a well defined quantity anymore since we are dealing with mixed states \cite{Phoenix1991b}. We have to rely on more sophisticated entanglement measures. One of the simplest such entanglement measures is the logarithmic negativity $E_N(\hat{\rho})$ \cite{Plenio2005,Plenio2005a,Wootters2001}. It is based on the negativity 
\begin{equation} \label{eq:Negativity}
N(\hat{\rho})  = \sum_i \frac{|\lambda_i| - \lambda_i}{2},
\end{equation}
where the $\lambda_i$ are the eigenvalues of the partial transpose of $\hat{\rho} = \sum\rho_{ijkl}\ket{i}\bra{j}\otimes\ket{k}\bra{l} $ over subsystem A defined as
\begin{equation} \label{eq:Transpose}
\hat{\rho}^{\Gamma_A}  = \sum_{i,j,k,l}\rho_{ijkl}\ket{j}\bra{i}\otimes\ket{k}\bra{l}.
\end{equation}
The logarithmic negativity can be calculated from the negativity by
\begin{equation} \label{eq:LogNeg}
E_N(\hat{\rho}) = \log_2(2N(\hat{\rho})+1).
\end{equation}
It can be shown that the result does not depend on the subsystem over which the transpose is taken \cite{Eisert1998,Plenio2005a}. Its value falls in the range $0 \leq E_N \leq 1$ where 0 is a pure state and 1 a maximally entangled state. $E_N$ is also an upper bound to the distillable entanglement $E_D$, a result to which we shall come back in our later discussion of teleportation.\\
\begin{figure}[t]
\begin{center}
\includegraphics[width = \columnwidth]{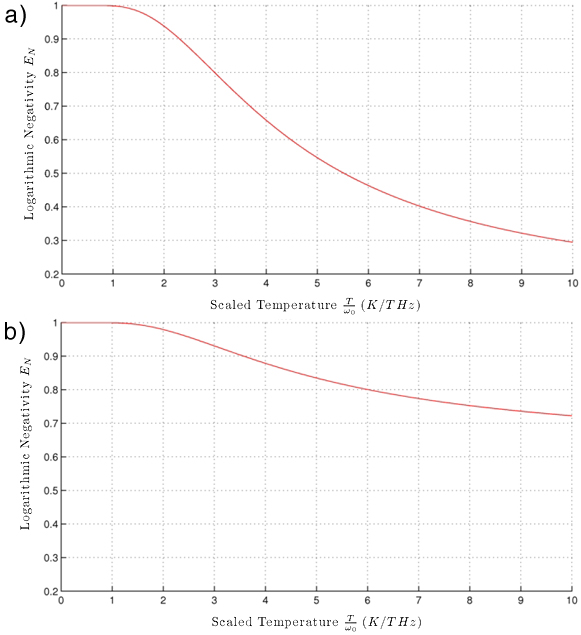}
\caption{Logarithmic negativity $E_N$ as a function of the scaled temperature $\frac{T}{\omega_0}$ for a) the Browne- Plenio scheme with an interaction time $\lambda\tau=0.01\pi$ and b) the scheme we propose with $\lambda\tau=0.288\pi$.}
\label{Negativity}
\end{center}
\end{figure}
After taking the partial transpose over cavity A, we numerically calculated the logarithmic negativity of the state $\hat{\rho}_{cav}$. Figure \ref{Negativity}a shows $E_N$ after an interaction time $\lambda\tau=0.01\pi$, i.e. very close to the ideal time $\lambda\tau \rightarrow 0$, plotted against the scaled temperature $\frac{T}{\omega_0}$. For low temperatures $\frac{T}{\omega_0} \lesssim 1 \frac{K}{THz}$ the logarithmic negativity is very flat and almost equal to 1 implying near-perfect entanglement. Only for temperatures higher than this does $E_N$ substantially fall off. Hence if we assume that our cavity operates in the microwave range of the spectrum with $\omega_0 \approx 10~GHz$, we can operate our cavities at temperatures of up to $T\approx100 K$ and still achieve very high levels of entanglement. We can also use these values to justify truncating the thermal states for $n>1$. For the highest temperatures shown in Figure \ref{Negativity}, $\frac{T}{\omega_0} = 10 \frac{K}{THz}$, we have according to \eqref{eq:nbar} $\bar{n} \approx 0.87$. Substitution into \eqref{eq:ThermalFieldCoeff} gives $P_0 \approx 0.53 > P_1 \approx 0.25 > P_2 \approx 0.11$. For the region we are particularly interested in, $\frac{T}{\omega_0} \lesssim 1 \frac{K}{THz}$, these results are even more dramatic, with $\bar{n} \approx 4.8\times10^{-4}$ and \begin{equation}
P_0 \approx 0.9995 \gg P_1 \approx 4.8\times10^{-4} \gg P_2 \approx 2.3\times10^{-7},
\end{equation}
which justifies the approximation made by omitting terms containing $P_2$ or higher.\\

\subsubsection{Our Scheme}
For the scheme we proposed, we analysed the case where the atom is thermally interacting with its environment and the cavities are in the state $\ket{1}_A\ket{0}_B$ as before, i.e. $\hat{\rho}(0) = \hat{\rho}_{Th}^A \otimes \ket{1}\bra{1}_A \otimes \ket{0}\bra{0}_B$ . Since $\hat{\rho}_{Th}^A$ only contains two terms, we do not have to make any approximations as in the previous case. Following the same procedure outlined above for the Browne-Plenio scheme, we arrive at the density matrix for the cavities after the atom has been measured in $\ket{g}$
\begin{equation} \label{eq:rhocav}
\hat{\rho}_{cav}  = \begin{pmatrix}
0 & 0 & 0 & 0 & 0 & 0\\
0 & P_g s_1^4 & -P_g c_1 s_1^2 & 0 & 0 & 0\\
0 & -P_g c_1 s_1^2 & P_g c_1^2 & 0 & 0 & 0\\
0 & 0 & 0 & P_e c_2^2 s_1^2 & P_e c_2 s_2 s_1 & 0\\
0 & 0 & 0 & P_e c_2 s_2 s_1 & P_e s_2^2 & 0\\
0 & 0 & 0 & 0 & 0 & 0
\end{pmatrix}
\end{equation}
in the basis $\{ \ket{00}, \ket{01}, \ket{10}, \ket{11}, \ket{20}, \ket{21}\}$, where we have introduced the shorthand notation
\begin{eqnarray}
c_n & = & \cos(\lambda\tau\sqrt{n}), \\
s_n & = & \sin(\lambda\tau\sqrt{n}).
\end{eqnarray}
This can now in fact be directly compared with the Bell state density matrix $\hat{\rho}_{\Psi^-} = \ket{\Psi^-}\bra{\Psi^-}$ in the basis $\{ \ket{00}, \ket{01}, \ket{10}, \ket{11}\}$.
\begin{equation}
\hat{\rho}_{\Psi^-}  = \begin{pmatrix}
0 & 0 & 0 & 0 \\
0 & +\frac{1}{2} & -\frac{1}{2} & 0 \\
0 & -\frac{1}{2} & +\frac{1}{2} & 0 \\
0 & 0 & 0 & 0
\end{pmatrix}
\end{equation}
We find that there are two conditions that have to be satisfied for our scheme to produce a Bell state. The first condition is $P_e\rightarrow0$, i.e. $T\rightarrow0$, which was to be expected since the ideal scheme assumes that the atom is in the ground state. The second condition is
\begin{equation} 
\sin^4(\lambda\tau) = \cos^2(\lambda\tau) = \cos(\lambda\tau)\sin^2(\lambda\tau)
\end{equation}
which has the real solutions
\begin{equation}  \label{eq:Solution}
\lambda\tau = 2\Bigl(\pi n \pm \tan^{-1}\sqrt{\sqrt{5}-2}\Bigr),
\end{equation}
with the lowest positive solution $\lambda\tau \approx 0.288\pi$, which is slightly longer than the ideal value of the thermally isolated scheme, $\lambda\tau = \frac{\pi}{4}$.\\
Further, we calculated the logarithmic negativity for this scheme in the way described above. The results are plotted in \ref{Negativity}b for an interaction time $\lambda\tau = 0.288\pi$. The results look remarkably similar to the results found for the Browne-Plenio scheme. The low temperature behaviour is equivalent for both, having a near perfect entanglement for $\frac{T}{\omega_0} \lesssim 1 \frac{K}{THz}$. Our scheme appears to be considerably less sensitive to increasing temperature and hence more reliable. However, a direct comparison is difficult since in one case the field modes and in the other the atom are allowed to interact thermally.

\section{Applications to Teleportation}
\begin{figure}[t]
\begin{center}
\includegraphics[width = \columnwidth]{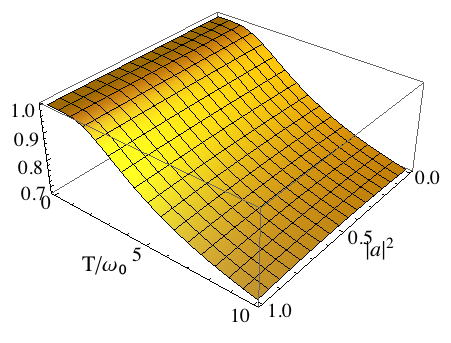}
\caption{Fidelities $F_a = F_d$ of the teleported state as a function of scaled temperature $\frac{T}{\omega_0}$ (in $\frac{K}{THz}$) and $|a|^2$. The interaction time is $\lambda\tau = 0.288\pi$, close to the ideal value.}
\label{Fidelity2}
\end{center}
\end{figure}

Finally, we want to analyse the possibility of applying our scheme to achieve quantum teleportation \cite{Bennett1993}. Let us assume that once we have created the entangled state $\hat{\rho}_{cav}$ we can transmit the two cavity modes, say cavity mode A to Alice and B to Bob, without the state decohering. We now want to make use of the entanglement that Alice and Bob share to transmit an unknown qubit $\ket{\psi}_C$
\begin{equation} \label{eq:Initial}
\ket{\psi}_C =  a\ket{0}_C + b\ket{1}_C
\end{equation}
that Alice possesses to Bob. We want to transmit the state using only local operations and classical communication (LOCC). This is the concept of quantum teleportation \cite{Plenio1998, Gerry2004, Preskill1998}.\\
Let us first assume that Alice and Bob share a perfect Bell state $\ket{\Psi^-}_{AB}$ and leave the generalisation to our imperfect state for later. The total three-qubit system is in the state
\begin{equation} \label{eq:Total}
\ket{\chi} =  \ket{\Psi^-}_{AB}\otimes\ket{\psi}_C
\end{equation}
By making full use of the Bell states \eqref{eq:Bell1} and \eqref{eq:Bell2} we can rewrite this state as
\begin{equation} \label{eq:Tele}
\begin{split}
\ket{\chi} = \frac{1}{2} \Bigl[ & \ket{\Psi^+}_{AC}\otimes(-a \ket{0}_B + b \ket{1}_B) \\
& \ket{\Psi^-}_{AC}\otimes(+a \ket{0}_B + b \ket{1}_B) \\
& \ket{\Phi^+}_{AC}\otimes(+a \ket{1}_B - b \ket{0}_B) \\
& \ket{\Phi^-}_{AC}\otimes(+a \ket{1}_B + b \ket{0}_B)
\Bigr].
\end{split}
\end{equation}
So far there was no action by Alice or Bob involved, only a change of basis in which we express our state. Now we can immediately see what happens when Alice makes a local measurement on her two-qubit state $AC$ in the Bell basis. The state that Bob possesses will collapse to one of four possible states, depending on the outcome of Alice's measurement. The results are
\begin{eqnarray}
\ket{\Psi^-}_{AC} & \rightarrow & +a \ket{0}_B + b \ket{1}_B, \\
\ket{\Phi^-}_{AC} & \rightarrow & +a \ket{1}_B + b \ket{0}_B, \\
\ket{\Phi^+}_{AC} & \rightarrow & +a \ket{1}_B - b \ket{0}_B, \\
\ket{\Psi^+}_{AC} & \rightarrow & -a \ket{0}_B + b \ket{1}_B,
\end{eqnarray}
where the notation $\ket{x} \rightarrow \ket{y}$ stands for Alice measuring $\ket{x}$ and Bob's state collapsing to $\ket{y}$. Now Alice can tell Bob via a classical communication channel what she measured, which will tell Bob the state of his qubit. If Alice measured $\ket{\Psi^-}_{AC}$ his qubit immediately takes the form $\ket{\chi}$, thus completing the teleportation process whithout Bob having to do anything. However, if Alice measures the states $\ket{\Phi^-}_{AC} $, $\ket{\Phi^+}_{AC}$ or $\ket{\Psi^+}_{AC}$, Bob can use the local operations $\hat{\sigma}_x$, $i\hat{\sigma}_y$ or $\hat{\sigma}_z$ respectively (the $\hat{\sigma}_i$ being the usual Pauli matrices) to transform his qubit into the desired state $\ket{\chi}$, thereby also completing the teleportation process \cite{Preskill1998}.\\
In the process, the entanglement that Alice and Bob shared is ``used up''. There are no correlations between Bob's qubit B and Alice's two qubits A or C anymore. However, A and C are now in a Bell state, but they are only locally available to Alice, so they cannot be used for further teleportation between Alice and Bob. It thus makes sense to speak of entanglement as a resource.\\
Now that we have worked out the basic teleportation protocol we can analyse what would happen if Alice and Bob do not share a perfect Bell state but the state $\hat{\rho}_{cav}$ given by equation \eqref{eq:rhocav}. The total state of the three-qubit system is described by the density operator
\begin{equation}
\hat{\rho}_{tot} = \hat{\rho}_{cav}\otimes\ket{\psi}\bra{\psi}_C.
\end{equation}
Alice now makes the same measurement in the Bell basis on her two qubits A an C as before. She thereby projects Bob's qubit into one of the following states
\begin{eqnarray}
\hat{\rho}_B^a & = & \braket{\Psi^-_{AC} | \hat{\rho}_{tot} | \Psi^-_{AC}},\\
\hat{\rho}_B^b & = & \braket{\Phi^-_{AC} | \hat{\rho}_{tot} | \Phi^-_{AC}},\\
\hat{\rho}_B^c & = & \braket{\Phi^+_{AC} | \hat{\rho}_{tot} | \Phi^+_{AC}},\\
\hat{\rho}_B^d & = & \braket{\Psi^+_{AC} | \hat{\rho}_{tot} | \Psi^+_{AC}}.
\end{eqnarray}
Bob then has to apply the same local transformations given by the Pauli matrices as in the ideal case described above. The fidelities of the final teleported states compared to the original state, after Bob has made the necessary  transformation, are given by
\begin{eqnarray}
F_a & = & |\braket{\psi | \hat{\rho}_B^a | \psi}|,\\
F_b & = & |\braket{\psi | \hat{\sigma}_x\hat{\rho}_B^b \hat{\sigma}_x| \psi}|,\\
F_c & = & |\braket{\psi | i\hat{\sigma}_y\hat{\rho}_B^ci\hat{\sigma}_y | \psi}|,\\
F_d & = & |\braket{\psi | \hat{\sigma}_z\hat{\rho}_B^d\hat{\sigma}_z | \psi}|.
\end{eqnarray}
We calculated the explicit forms of these fidelities and found
\begin{equation}\label{eq:Fid1}
\begin{split}
F_a & = F_d \\
& = \frac{P_g(|a|^4c_1^2 + 2|a|^2|b|^2c_1s_1^2 + |b|^4s_1^4) + P_e|a|^2|b|^2c_2^2s_1^2}{|a|^2c_1^2 + |b|^2s_1^4}
\end{split}
\end{equation}
and
\begin{equation}\label{eq:Fid2}
\begin{split}
F_b & = F_c \\
& = \frac{P_g(|a|^4s_1^4 + 2|a|^2|b|^2c_1s_1^2 + |b|^4c_1^2) + P_e|a|^2|b|^2c_2^2s_1^2}{|a|^2s_1^4 + |b|^2c_1^2}.
\end{split}
\end{equation}
We can eliminate one variable by insisting that the state $\ket{\psi}$ is normalised, implying $|a|^2 + |b|^2 = 1$. The fidelities are plotted as functions of temperature and $|a|^2$ in Figure \ref{Fidelity2} for the near-ideal interaction time $\lambda\tau = 0.288\pi$. The plot explicitly shows $F_a = F_d$, but the other plot for $F_b = F_c$ is essentially indistinguishable and was therefore omitted. In fact, a numerical analysis shows that $F_a=F_b = F_c = F_d$ for $\lambda\tau$ going to one of the values in equation \eqref{eq:Solution}.
\\
Surprisingly, we see that now the fidelity of the teleported qubit has a direct dependence on the precise nature of the original qubit, through the $|a|^2$ dependence in equations \eqref{eq:Fid1} and \eqref{eq:Fid1}, unlike in the ideal condition protocol, which is completely independent of the explicit form of $\ket{\psi}$. Yet, the figure shows that this effect is barely noticeable and converges to the correct value $F=1$ as $T\rightarrow 0$ for all $|a|^2$. We also note that the fidelity drops with increasing temperature, in line with the results found for the thermal states in section IV.D. In fact, the temperature dependence of the fidelity shows a remarkable resemblance to the temperature dependence of the logarithmic negativity shown in Figure \ref{Negativity}. As was to be expected, if we want to achieve perfect teleportation using only one entangled qubit pair produced by our scheme, we have to go to infinitesimally low temperatures.\\
However, there exist more realistic ways of realising a perfect teleportation protocol. It can be shown \cite{Vedral1998,Plenio2005a} that given a certain number $N$ of non-maximally entangled qubit pairs such as our state \eqref{eq:rhocav}, it is possible to ``purify'' the entanglement, producing $M<N$ maximally entangled states, using local operations only. The crucial quantity limiting the efficiency of this purification process is the distillable entanglement $E_D$ briefly mentioned in the discussion of the logarithmic negativity. The condition limiting the efficiency is $M \leq N E_D$. We can turn this into an inequality involving the logarithmic negativity calculated earlier, since as mentioned above the logarithmic negativity is an upper bound on the distillable entanglement, $E_N \geq E_D$. This implies
\begin{equation}
N \geq \frac{M}{E_N(\hat{\rho})}.
\end{equation}
Hence if we want to use our scheme to provide $M$ maximally entangled qubit-pairs that can then each be used to teleport one qubit, we need at least $N_{min}=\frac{M}{E_N(\hat{\rho}_{cav})}$ copies of the state $\hat{\rho}_{cav}$ shared between Alice and Bob. Note that this only allows us to determine a lower limit on the necessary number of copies. There is no guarantee that this number will be sufficient to actually achieve the desired number of maximally entangled states through the purification process. Nevertheless, these considerations at least qualitatively show that it is possible to use our scheme to generate entangled states that can be sent to two parties Alice and Bob, which can carry out entanglement purification procedures, and then use the resulting states for teleportation with unit fidelity. Considering again the results in Figure \ref{Negativity} we can see that even at relatively high temperatures it is possible to provide states for the teleportation protocol, as long as we are willing to sacrifice efficiency. Quantitatively we can only place a lower bound on the required number of copies needed. Under real conditions this is likely to be increased even further by the fact that the states could undergo further decoherence as they are distributed to Alice and Bob.\\

\section{Conclusions}
We have derived the Rabi model Hamiltonian and justified its simplification under the rotating wave approximation into the Jaynes-Cummings model Hamiltonian. These results lead us to show that the JCM predicts two distinct interaction processes between the atom and the field. Near resonance the two subsystems can exchange energy, whereas for large detuning only the phases of the two subsystems are affected. In section III.B we showed how both of these could be used to entangle an atom-field system.\\
We then extended the scheme developed by Browne and Plenio \citep{Browne2002a} for entangling two cavities by proposing a slightly modified scheme, promising a considerably higher success rate and a fidelity only limited by the (im)precision of the experimental setup.\\
We further added realism to both schemes by allowing the systems to thermally interact with the environment and showed that in this case the fidelities are monotonously decreasing functions of temperature.\\
Finally we considered the applicability of our scheme to generating states that can be used for quantum teleportation. We showed that the fidelity of the teleported state also monotonously decreases with temperature. Nevertheless, perfect teleportation is possible if we add a purification scheme to the protocol, which would in turn require a larger number of shared states to be generated and distributed. We were able to place a lower bound on this number related to the logarithmic negativity of the state. A possible future step would be to treat the cavities not as perfect mirrors, but as beamsplitters with a small but finite transmittance and investigate how this would affect the fidelity of the final state.\\
All these questions will be of crucial importance as more and more quantum information protocols, such as quantum teleportation, are experimentally realised.

\section*{Acknowledgments}
I thank Dr. Tommaso Tufarelli for his help and support, as well as Billy Hesmondhalgh for his collaboration on this project.

\bibliographystyle{phaip}

\nocite{Solano2002}

\bibliography{JCM_MaxFrenzel}

\end{document}